\documentclass[twocolumn,showpacs,preprintnumbers,amsmath,amssymb]{revtex4}
\usepackage{graphicx}% Include figure files
\usepackage{dcolumn}% Align table columns on decimal point
\usepackage{bm}% bold math

\begin{document}

\title{$\alpha$ particle preformation in heavy nuclei and penetration probability}

\author{H.F. Zhang$^{1}$}

\author{G. Royer$^{2}$}
\email{royer@subatech.in2p3.fr}

\affiliation{\footnotesize$^1$ School of Nuclear Science and
Technology, Lanzhou University, Lanzhou 730000, People's Republic
of China}

\affiliation{\footnotesize $^2$Laboratoire Subatech, UMR:
IN2P3/CNRS-Universit\'e-Ecole des Mines, 44307 Nantes Cedex 03, France}

\date{\today}

\begin{abstract}
The $\alpha$ particle preformation in
the even-even nuclei from $^{108}$Te to $^{294}$118  
and the penetration probability have been studied. The isotopes from Pb to
U have been firstly investigated since the experimental data allow to extract the microscopic
features for each element. The
assault frequency has been estimated using classical methods and the
penetration probability from tunneling through the
 Generalized Liquid Drop Model (GLDM) potential barrier. The preformation factor has been extracted from
experimental $\alpha$ decay energies and half-lives. The shell closure effects play the key role in
the $\alpha$ preformation. The more the nucleon number is close to the magic numbers, the more the formation of $\alpha$ cluster is difficult inside the mother nucleus. The penetration probabilities reflect that 126 is a neutron magic number. The penetration probability range is very
large compared to that of the preformation factor. The penetration
probability determines mainly the $\alpha$
decay half-life while the preformation factor allows to obtain information on the nuclear structure. The study has been extended to the newly observed heaviest nuclei.

\end{abstract}
\pacs{23.60.+e, 21.10.Jx, 27.90.+b}

\maketitle

\section{Introduction}
The $\alpha$ emission is one of the decay channels of the heavy
nuclei. Measurements on the $\alpha$ decay can provide reliable information on the nuclear structure such
as the ground-state energy, the ground-state half-life, the nuclear spin
and parity, the nuclear deformation, the nuclear clustering, the shell effects and the nuclear interaction \cite{Re87,Ho91,Fi96,Lo98, Gar00,Au03,Gi03,Ho03,He04,Ga04,Se06,Lep07}.
 Measurements on the $\alpha$ decay are also used to
identify new nuclides or new heavy elements since the $\alpha$ decay allows
 to extract clear and reliable information on the parent
nucleus \cite{Og05,Hof07,Mo07}. Recently, the interest in the $\alpha$ decay 
has been renewed because of the
development of radioactive beams and new detector technology under
low temperature. 

The $\alpha$ decay process was first explained by Gamow and
by Condon and Gurney in the 1920s \cite{Ga28,Co28} as a
quantum-tunneling effect and is one of the first examples proving the need to use the quantum mechanics to describe 
the nuclear phenomena and its correctness. Later on, theoretical
calculations were performed to predict the absolute $\alpha$ decay
width, to extract nuclear structure information, and to pursue a
microscopic understanding of the $\alpha$-decay phenomenon. These
studies are based on various theoretical models such as the shell
model, the fissionlike model, and the cluster model
\cite{Po83,Br92,Va92,Bu93,Roy00,Gu02,Du02,Ba03,Zh06,Mo06,Re06,Po06,Pe07}.
In the $\alpha$ cluster model, the decay constant $\lambda$ is
 the product of three terms: the cluster preformation
probability $P_{0}$, the assault frequency $\nu_{0}$ and the
barrier penetrability P. The $\alpha$
particle emission is a quantum tunnelling through the potential
barrier leading from the mother nucleus to the two emitted
fragments: the $\alpha$ particle and the daughter nucleus. The penetrability 
can be described successfully using the WKB approximation. The most
important problem of the $\alpha$ decay is how to estimate the
preformation probability $P_{0}$ that the $\alpha$ particle exists as a recognizable
entity inside the nucleus before its emission. It is very dependent
on the structure of the states of the mother and daughter nuclei
and is a measure of the degree of resemblance between the initial state of the mother nucleus and 
the final state of the daughter nucleus plus the $\alpha$
particle. Within an asymmetric fission picture the
preformation factor P$_{0}$ has been taken as 1 in previous studies
and that allows to reproduce the experimental $\alpha$ decay half-lives \cite{Roy00,Zh06} 
when the experimental $Q_{\alpha}$ values are used. There
are, however, still small differences between the calculated and
experimental values and these discrepancies may be used to determine the $\alpha$-preformation probability.
The $\alpha$ preformation factor is very important from the viewpoint of the nuclear structure. Numerous 
studies of the $\alpha$ decay have been concentrated on this
problem \cite{Ma60,Ar74,Fl76,Ga83,Du86,Re88,Ro98,Ha00}. There are also
several approaches to calculate the formation amplitude : the
shell model, the BCS method \cite{del96}, the hybrid (shell
model+$\alpha$-cluster) model \cite{var92} etc. The effect of
continuum states on the $\alpha$ decay is known to be very large
\cite{del96}. One needs therefore very large shell model basis to
obtain the experimental values of the $\alpha$-preformation
probability. The hybrid model \cite{var92},
which treats a large shell model basis up to the continuum states
through the wavefunction of the spatially localized $\alpha$
cluster, explains well the experimental decay width. In these
calculations the wave function is necessary, and it is not easy
to extend the approaches for nuclei including more nucleons
outside the double-magic core. 

In this contribution, the
preformation factor is extracted from the experimental
$\alpha$ decay half-life $T_{\alpha}$ and the penetration
probability is obtained from the WKB approximation. The potential
barrier is determined within a generalized liquid drop model
(GLDM) including the proximity effects between the $\alpha$ particle
and the daughter nucleus and adjusted to reproduce the
experimental Q$_{\alpha}$ \cite{Roy00,Zh06}.

The paper is organized as follows. In Sec. 2, the methods for calculating the assault
frequencies, the penetrability of the even-even nuclei and the preformation factor are presented. The
calculated results are shown and discussed in Sec. 3. In the last
section, the conclusions are given and future works are
proposed.

\section{Methods}
The $\alpha$ decay constant is defined as
\begin{equation}\label{lamda1}
\lambda=P_{0}\nu_{0}P.
\end{equation}
Imagining the $\alpha$ particle moving
back and forth inside the nucleus with a velocity $v ~=~\sqrt{\frac{2E_{\alpha}}{M}} ~$, it presents
itself at the barrier with a frequency :
\begin{eqnarray}\label{mu}
\nu_{0} ~=~\left(\frac{1}{2R}\sqrt{\frac{2E_{\alpha}}{M}}\right)~.
\label{orig_nu}
\end{eqnarray}
R is the radius of the parent nucleus given by
\begin{equation}\label{radii}
R_i=(1.28A_i^{1/3}-0.76+0.8A_i^{-1/3}) \ \textrm{fm},
\end{equation}
and E$_{\alpha}$ is the energy of the alpha particle, corrected for
recoil; $M$ being its mass. 

The penetration probability P is calculated within the WKB approximation. The potential barrier governing the $\alpha$ emission is determined within the GLDM, including the volume, surface, Coulomb
and proximity energies \cite{Roy85}:
\begin{equation}\label{etot}
E ~= ~E_{V}+E_{S}+E_{C}+E_{\text{Prox}} ~.
\end{equation}
When the nuclei are separated:
\begin{equation}\label{ev}
E_{V}=-15.494\left \lbrack (1-1.8I_1^2)A_1+(1-1.8I_2^2)A_2\right
\rbrack \ \textrm{MeV},
\end{equation}
\begin{equation}\label{es}
E_{S}=17.9439\left
\lbrack(1-2.6I_1^2)A_1^{2/3}+(1-2.6I_2^2)A_2^{2/3} \right \rbrack
\ \textrm{MeV},
\end{equation}
\begin{equation}\label{ec}
E_{C}=0.6e^2Z_1^2/R_1+0.6e^2Z_2^2/R_2+e^2Z_1Z_2/r,
\end{equation}
where $A_i$, $Z_i$, $R_i$ and $I_i$ are the mass numbers, charge
numbers, radii and relative neutron excesses of the two nuclei. $r$
is the distance between the mass centres. The radii $R_{i}$ are
given by Eq. (\ref{radii}).

For one-body shapes, the surface and Coulomb energies are defined
as \cite{Roy85,Roy00}:
\begin{equation}\label{esone}
E_{S}=17.9439(1-2.6I^2)A^{2/3}(S/4\pi R_0^2) \ \textrm{MeV},
\end{equation}
\begin{equation}\label{econe}
E_{C}=0.6e^2(Z^2/R_0) \times 0.5\int
(V(\theta)/V_0)(R(\theta)/R_0)^3 \sin \theta d \theta.
\end{equation}
$S$ is the surface of the one-body deformed nucleus. $V(\theta )$
is the electrostatic potential at the surface and $V_0$ the
surface potential of the sphere.

The surface energy results from the effects of the surface tension
forces in a half space. When there are nucleons in regard in a
neck or a gap between separated fragments an additional term
called proximity energy must be added to take into account the
effects of the nuclear forces between the close surfaces. This
term is essential to describe smoothly the one-body to two-body
transition and to obtain reasonable fusion barrier heights. It
moves the barrier top to an external position and strongly
decreases the pure Coulomb barrier.
\begin{equation}
E_{\text{Prox}}(r)=2\gamma \int _{h_{\text{min}}}
^{h_{\text{max}}} \Phi \left \lbrack D(r,h)/b\right \rbrack 2 \pi
hdh,
\end{equation}
where $h$ is the distance varying from the neck radius or zero to
the height of the neck border. $D$ is the distance between the
surfaces in regard and $b=0.99$~fm the surface width. $\Phi$ is
the proximity function. The surface
parameter $\gamma$ is the geometric mean between the surface
parameters of the two fragments. The
combination of the GLDM and of a quasi-molecular shape sequence
has allowed to reproduce the fusion barrier heights and radii, the
fission and the $\alpha$ and cluster radioactivity data.

The barrier penetrability $P$ is calculated within the action
integral
\begin{equation}\label{penetrability}
    P ~= ~exp[-\frac{2}{\hbar}\int_{R_{\text{in}}}^{R_{\text{out}}}\sqrt{2B(r)(E(r)-E(sphere))}] ~.
\end{equation}
The deformation energy (relatively to the sphere energy) is small till the
rupture point between the fragments \cite{Roy00} and the two
following approximations may be
used : $R_{\text{in}}=R_{d}+R_{\alpha}$ and  $B(r)=\mu$ where $\mu$
is the reduced mass. $R_{out}$ is simply
e$^{2}$Z$_{d}$Z$_{\alpha}$/Q$_{\alpha}$.

The decay constant may be deduced from the
experimental $\alpha$ decay half-life $T_{\alpha}$ by
\begin{equation}\label{lamda2}
\lambda ~ = ~ \frac{ln2}{T_{\alpha}}.
\end{equation}
The
preformation factor $P_{0}$ of an $\alpha$ cluster inside the mother
nucleus can be estimated inserting Eq.(\ref{mu}), Eq.(\ref{penetrability}) and
Eq.(\ref{lamda2}) in Eq.(\ref{lamda1}).

\section{Results and discussions}

The values of the decay constant $\lambda$, the velocity, the assault frequency $\nu_{0}$, the penetrability P and the preformation factor 
P$_{0}$ are given for the even-even Pb, Po, Rn, Ra, Th and U isotopes in the tables 1 and 2. 

\begin{figure*}[t]
%\centering
\includegraphics[scale=0.9]{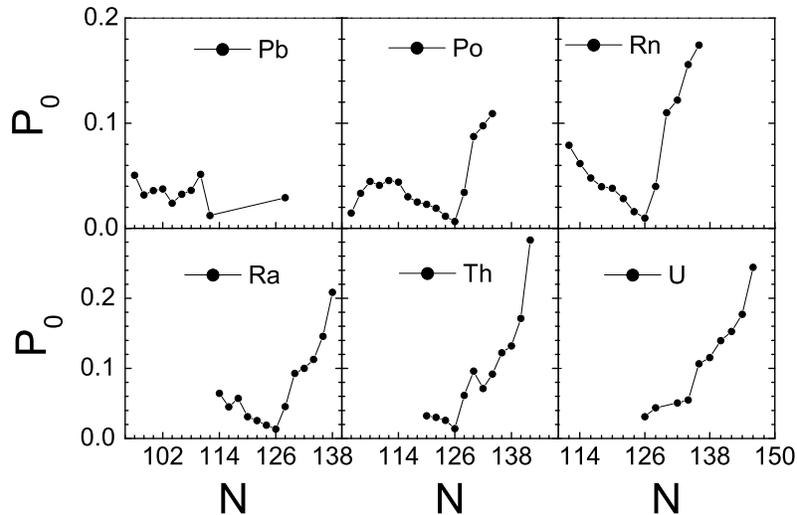}
\caption{Extracted preformation factors for the even-even isotopes of
Pb, Po, Rn, Ra, Th, and U.}
\label{Fig1}
\end{figure*}

\begin{figure*}[t]
%\centering
\includegraphics[scale=0.8]{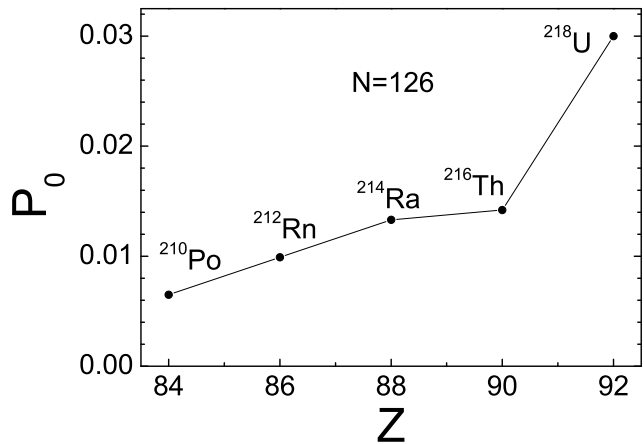}
\caption{Extracted preformation factors for the N=126 isotones of Po, Rn, Ra, Th, and U.}
\label{Fig2}
\end{figure*}

\begin{figure*}[t]
%\centering
\includegraphics[scale=0.8]{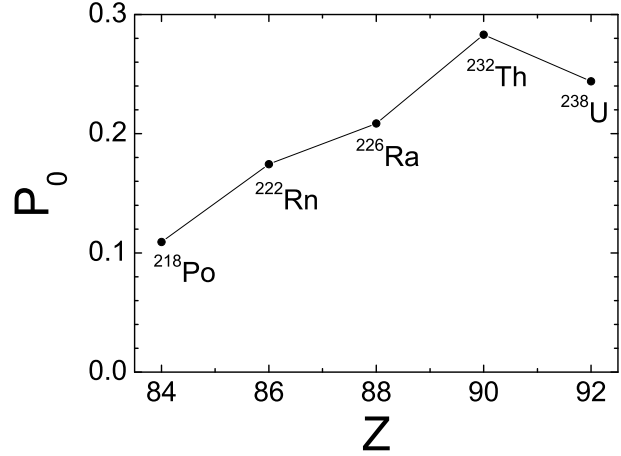}
\caption{Extracted preformation factors for $^{218}$Po,
$^{222}$Rn, $^{226}$Ra, $^{232}$Th, and $^{238}$U.}
\label{Fig3}
\end{figure*}

In the simple picture of an $\alpha$ particle moving with a
kinetic energy E$_{\alpha}$ inside the nucleus, the order of magnitude of the $\alpha$ velocity $v$ is
1.0$\times$10$^{22}$ fm/s for all considered nuclei. The order of
magnitude of the assault frequency $\nu_{0}$ calculated by
Eq.(\ref{lamda1}) is 1.0$\times$10$^{21}s^{-1}$ both in the table 1 and the 
table 2. The variations of $v$ and $\nu_{0}$ are small.
 In the previous study relative to the $\alpha$ decay half-lives using the GLDM \cite{Roy00}, the
assault frequency $\nu_{0}$ is fixed as 1.0$\times$10$^{20}$,
which is smaller by one order of magnitude than the value obtained in the
present calculation, meaning that an averaged preformation factor of
0.1 has been included in the previous $\alpha$ decay study. So
the previous calculations are consistent quantitatively with the
experimental data in general. A fixed value of the assault
frequency cannot describe the detailed features of the nuclear
structure and the last
column of the tables 1 and 2 displays the extracted preformation factor P$_{0}$ when the Eq. (\ref{mu}) is assumed.
To study the correlation between the preformation factors and the
structure properties, the preformation
factors of Pb, Po, Rn, Ra,
Th, and U isotopes are given as a function of the neutron number N in Fig.\ref{Fig1}. The
preformation factors of the isotopes generally decrease with
increasing neutron number up to the spherical shell closure N=126,
where the minimum of P$_{0}$ occurs, and then they increase
quickly with the neutron number. The values of the preformation
factors is always minimum for the Po, Rn, Ra, Th, and U
isotopes when N=126. The preformation
factors of N=126 isotones are shown in Fig. \ref{Fig2}. The
preformation factors increase when the proton number goes away from the magic number Z=82. The maximum
values of the preformation factor for the five elements, which correspond to the most
neutron-rich nucleus from Po to U isotopes respectively, are
presented in Fig.\ref{Fig3}. In general, the values increase with the proton number. From $^{232}$Th to $^{238}$U,
the value of the preformation factor decreases apparently, even if 
the proton number Z is far from the magic number Z=82. The neutron
number of $^{238}$U (N=146) is closer to the submagic neutron number N=152
than that of the nucleus $^{232}$Th (N=142); so the preformation factor
of $^{232}$Th is larger than that of the nucleus $^{238}$U. As a conclusion 
the shell closure effects play the key
role in  the $\alpha$ formation mechanism. The more the nucleon number 
is close to the magic numbers, the more the formation of the $\alpha$ cluster 
inside the mother nucleus is difficult.

\begin{figure*}[t]
%\centering
\includegraphics[scale=0.9]{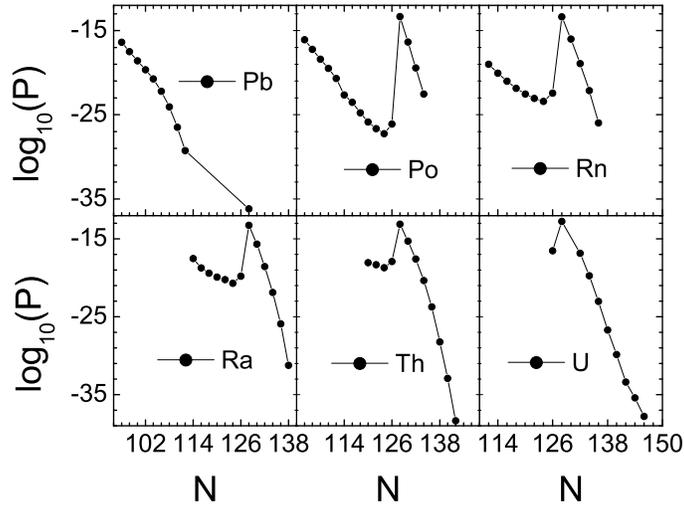}
\caption{Logarithms of the penetrability for the even-even isotopes of Pb,
Po, Rn, Ra, Th, and U.}
\label{Fig4}
\end{figure*}

The ground-state spin and parity of even-even nuclei is 0$^{+}$ and
the $\alpha$ decay of even-even nuclei mainly proceeds to the
ground state of the daughter nucleus. Though the parent nucleus
may also decay to the excited states of the daughter nucleus, the
probability is very small and it can be neglected
for a systematic study of penetrability. The
calculated penetrabilities are given in the seventh column in the tables 1 and 2. The 
values of the penetrability vary in a wide range
from 10$^{-39}$ to 10$^{-14}$ for the even-even isotopes of Pb,
Po, Rn, Ra, Th, and U, implying that the penetrability plays the key
role to determine the $\alpha$ decay half-lives. The values of log$_{10}$(P) are plotted in Fig. \ref{Fig4}
for the even-even Pb, Po, Rn, Ra, Th, and U isotopes. They decrease with the increasing neutron
number N for the Pb, Po, Rn, Ra, and Th isotopes before the magic
number N=126. For the U isotopes, the experimental data of
Q$_{\alpha}$ and T$_{\alpha}$ are not supplied. The penetrability
values increase sharply after N=126 and reach
the maximum at N=128, then they decrease again quickly for
Po, Rn, Ra, Th, and U isotopes. For the Pb isotopes the experimental
data of Q$_{\alpha}$ and T$_{\alpha}$ are not supplied after
N=128. For the nuclei with two neutrons outside the closed
shell, the $\alpha$ particle emission is easier than that of the other
nuclei of the same isotopes. The closed shell structures play also the
key role for the penetrability mechanism. The more the nucleon number 
is far away from the nucleon magic numbers, the more the penetrability of the
$\alpha$ cluster from mother nucleus becomes larger.

\begin{figure*}[t]
%\centering
\includegraphics[scale=0.9]{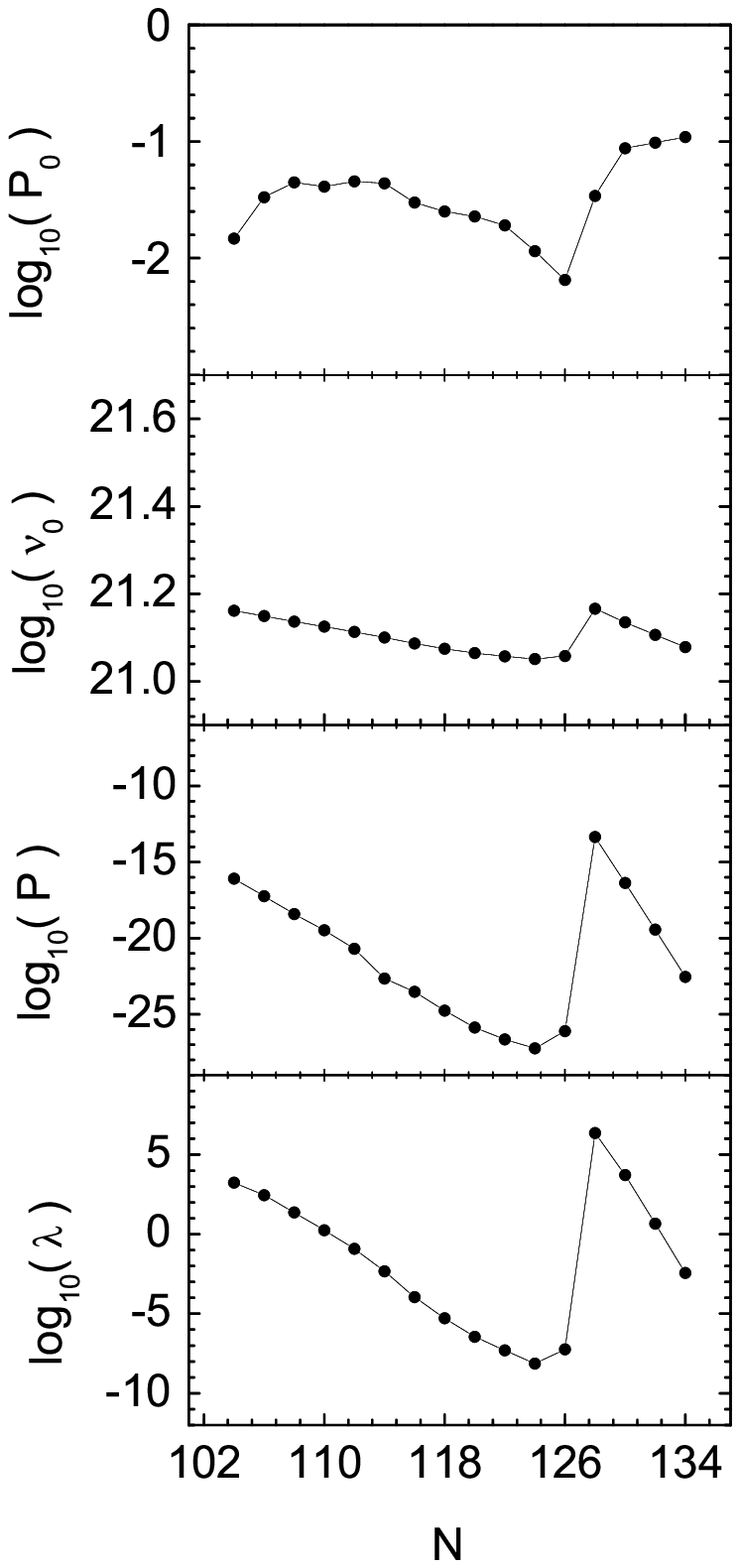}
\caption{Characteristics of the $\alpha$ formation and penetration for
the even-even Po isotopes.}
\label{Fig5}
\end{figure*}

The values of the $\alpha$ decay constant $\lambda$ deduced from the
experimental half-lives are shown in the fourth column of the tables 1
and 2 for the even-even isotopes of Pb, Po, Rn, Ra, Th, and U.
In order to clarify the relations between the decay constant $\lambda$,
the preformation factor P$_{0}$, the assault frequency $\nu_{0}$ and the penetrability P, the values of
log$_{10}$(P$_{0}$), log$_{10}$($\nu_{0}$), log$_{10}$(P), and
log$_{10}(\lambda)$ are displayed as functions of the neutron number N for Po isotopes in the 
Fig. \ref{Fig5}. The curve shape of log$_{10}(\lambda)$
is the same as that of log$_{10}$(P), confirming that the $\alpha$ decay
half-lives are mainly decided by
the penetrability. log$_{10}$($\nu_{0}$)
decreases smoothly before N=126, increases from N=126 to N=128, and
decreases again with the increasing of the neutron number, indicating
that the assault frequency of $\alpha$ particle can also reflect the
shell effects, even if log$_{10}$(P$_{0}$) exhibits the
microscopic nuclear structure more clearly.

\begin{table*}[h]
\label{table1} \caption{Characteristics of the $\alpha$ formation and
penetration for the even-even Pb, Po and Rn isotopes. The first column
indicates the mother nucleus. The second and third columns
correspond, respectively, to the experimental Q$_{\alpha}$ and
log$_{10}$[T$_{\alpha}$(s)]. The fourth column is the decay
constant $\lambda$. The fifth and
sixth columns correspond, respectively, to the velocity and
frequency of the $\alpha$ cluster inside the mother nucleus. The
seventh column is the penetration probability and the last column gives the preformation
factor extracted from Eq. (\ref{lamda1}) and the data
of this table.}
\begin{ruledtabular}
\begin{tabular}{llllllll}
$Nuclei$& Q$_{\alpha}$ [MeV]&log$_{10}$(T$_{\alpha}$)&$\lambda [ s^{-1} ]$ &$v [ fm/s ]$&$\nu_{0} [ s^{-1} ]$ & P &P$_{0}$\\
\hline
%&&&&&&& \\
$^{178}_{82}$Pb & 7.790  & -3.64 & 3.012 $\times 10^{3}$ & 1.913 $\times 10^{22}$&  1.453$\times 10^{21}$ & 4.097 $\times 10^{-17}$& 0.0506 \\
%&&&&&& \\
$^{180}_{82}$Pb & 7.415  & -2.30 & 1.386 $\times 10^{2}$ & 1.816 $\times 10^{22}$&  1.415$\times 10^{21}$ & 3.098 $\times 10^{-18}$ & 0.0317 \\
%&&&&&& \\
$^{182}_{82}$Pb & 7.080  & -1.26 & 1.261 $\times 10^{1}$ & 1.823 $\times 10^{22}$&  1.347$\times 10^{21}$ & 2.558 $\times 10^{-19}$ & 0.0359 \\
%&&&&&& \\
$^{184}_{82}$Pb & 6.774  & -0.21 & 1.137 $\times 10^{0}$ & 1.784 $\times 10^{22}$&  1.339$\times 10^{21}$ & 2.262 $\times 10^{-20}$ & 0.0375 \\
%&&&&&& \\
$^{186}_{82}$Pb & 6.470  &  1.08 & 5.779 $\times 10^{-2}$& 1.743 $\times 10^{22}$&  1.303$\times 10^{21}$ & 1.845 $\times 10^{-21}$ & 0.0240 \\
%&&&&&& \\
$^{188}_{82}$Pb & 6.109  &  2.43 & 2.569 $\times 10^{-3}$& 1.694 $\times 10^{22}$&  1.262$\times 10^{21}$ & 6.264 $\times 10^{-23}$ & 0.0325 \\
%&&&&&& \\
$^{190}_{82}$Pb & 5.697  &  4.26 & 3.853 $\times 10^{-5}$& 1.636 $\times 10^{22}$&  1.214$\times 10^{21}$ & 8.784 $\times 10^{-25}$ & 0.0361 \\
%&&&&&& \\
$^{192}_{82}$Pb & 5.221  &  6.56 & 1.927 $\times 10^{-7}$& 1.566 $\times 10^{22}$&  1.157$\times 10^{21}$ & 3.237 $\times 10^{-27}$ & 0.0514 \\
%&&&&&& \\
$^{194}_{82}$Pb & 4.738  &  9.99 & 7.077$\times 10^{-11}$& 1.491 $\times 10^{22}$&  1.098$\times 10^{21}$ & 5.278 $\times 10^{-30}$ & 0.0122 \\
%&&&&&& \\
$^{210}_{82}$Pb & 3.790  & 16.57 & 1.866$\times 10^{-17}$& 1.334 $\times 10^{22}$&  0.959$\times 10^{21}$ & 6.676 $\times 10^{-37}$ & 0.0293 \\
%&&&&&& \\
$^{188}_{84}$Po & 8.087  & -3.40 & 1.733 $\times 10^{3}$ & 1.950 $\times 10^{22}$&  1.452$\times 10^{21}$ & 8.129 $\times 10^{-17}$ & 0.0147 \\
%&&&&&& \\
$^{190}_{84}$Po & 7.693  & -2.60 & 2.772 $\times 10^{2}$ & 1.902 $\times 10^{22}$&  1.411$\times 10^{21}$ & 5.907 $\times 10^{-18}$ & 0.0333 \\
%&&&&&& \\
$^{192}_{84}$Po & 7.319  & -1.54 & 2.392 $\times 10^{1}$ & 1.855 $\times 10^{22}$&  1.371$\times 10^{21}$ & 3.909 $\times 10^{-19}$ & 0.0446 \\
%&&&&&& \\
$^{194}_{84}$Po & 6.990  & -0.41 & 1.782 $\times 10^{0}$ & 1.813 $\times 10^{22}$&  1.335$\times 10^{21}$ & 3.249 $\times 10^{-20}$ & 0.0410 \\
%&&&&&& \\
$^{196}_{84}$Po & 6.660  & 0.76  & 1.205 $\times 10^{-1}$& 1.770 $\times 10^{22}$&  1.299$\times 10^{21}$ & 2.041 $\times 10^{-21}$ & 0.0455 \\
%&&&&&& \\
$^{198}_{84}$Po & 6.310  & 2.18  & 4.580 $\times 10^{-3}$& 1.722 $\times 10^{22}$&  1.259$\times 10^{21}$ & 8.283 $\times 10^{-23}$ & 0.0439 \\
%&&&&&& \\
$^{200}_{84}$Po & 5.980  & 3.79  & 1.124 $\times 10^{-4}$& 1.677 $\times 10^{22}$&  1.222$\times 10^{21}$ & 3.066 $\times 10^{-24}$ & 0.0300 \\
%&&&&&& \\
$^{202}_{84}$Po & 5.700  & 5.13  & 5.138 $\times 10^{-6}$& 1.637 $\times 10^{22}$&  1.188$\times 10^{21}$ & 1.721 $\times 10^{-25}$ & 0.0251 \\
%&&&&&& \\
$^{204}_{84}$Po & 5.480  & 6.28  & 3.638 $\times 10^{-7}$& 1.605 $\times 10^{22}$&  1.161$\times 10^{21}$ & 1.376 $\times 10^{-26}$ & 0.0228 \\
%&&&&&& \\
$^{206}_{84}$Po & 5.330  & 7.15  & 4.907 $\times 10^{-8}$& 1.583 $\times 10^{22}$&  1.141$\times 10^{21}$ & 2.254 $\times 10^{-27}$ & 0.0191 \\
%&&&&&& \\
$^{208}_{84}$Po & 5.220  & 7.97  & 7.427 $\times 10^{-9}$& 1.567 $\times 10^{22}$&  1.126$\times 10^{21}$ & 5.727 $\times 10^{-28}$ & 0.0115 \\
%&&&&&& \\
$^{210}_{84}$Po & 5.407  & 7.08  & 5.765 $\times 10^{-8}$& 1.596 $\times 10^{22}$&  1.142$\times 10^{21}$ & 7.615 $\times 10^{-27}$ & 0.0065 \\
%&&&&&& \\
$^{212}_{84}$Po & 8.950  & -6.52 & 2.295 $\times 10^{6}$ & 2.054 $\times 10^{22}$&  1.466$\times 10^{21}$ & 4.598 $\times 10^{-14}$ & 0.0341 \\
%&&&&&& \\
$^{214}_{84}$Po & 7.830  & -3.87 & 5.138 $\times 10^{3}$ & 1.921 $\times 10^{22}$&  1.366$\times 10^{21}$ & 4.309 $\times 10^{-17}$ & 0.0873 \\
%&&&&&& \\
$^{216}_{84}$Po & 6.900  & -0.82 & 4.580 $\times 10^{0}$ & 1.803 $\times 10^{22}$&  1.278$\times 10^{21}$ & 3.670 $\times 10^{-20}$ & 0.0976 \\
%&&&&&& \\
$^{218}_{84}$Po & 6.110  & 2.27  & 3.722 $\times 10^{-3}$& 1.696 $\times 10^{22}$&  1.199$\times 10^{21}$ & 2.844 $\times 10^{-23}$ & 0.1092 \\
%&&&&&& \\
$^{198}_{86}$Rn & 7.349  & -1.19 & 1.066 $\times 10^{1}$ & 1.859 $\times 10^{22}$&  1.360$\times 10^{21}$ & 9.928 $\times 10^{-20}$ & 0.0790 \\
%&&&&&& \\
$^{200}_{86}$Rn & 7.040  &  0.0  & 6.931 $\times 10^{-1}$& 1.820 $\times 10^{22}$&  1.326$\times 10^{21}$ & 8.491 $\times 10^{-21}$ & 0.0616 \\
%&&&&&& \\
$^{202}_{86}$Rn & 6.770  & 1.06  & 6.037 $\times 10^{-2}$& 1.785 $\times 10^{22}$&  1.296$\times 10^{21}$ & 9.713 $\times 10^{-22}$ & 0.0480 \\
%&&&&&& \\
$^{204}_{86}$Rn & 6.550  & 2.00  & 6.931 $\times 10^{-3}$& 1.755 $\times 10^{22}$&  1.270$\times 10^{21}$ & 1.375 $\times 10^{-22}$ & 0.0397 \\
%&&&&&& \\
$^{206}_{86}$Rn & 6.380  & 2.71  & 1.352 $\times 10^{-3}$& 1.733 $\times 10^{22}$&  1.249$\times 10^{21}$ & 2.830 $\times 10^{-23}$ & 0.0382 \\
%&&&&&& \\
$^{208}_{86}$Rn & 6.260  & 3.34  & 3.168 $\times 10^{-4}$& 1.716 $\times 10^{22}$&  1.233$\times 10^{21}$ & 9.051 $\times 10^{-24}$ & 0.0284 \\
%&&&&&& \\
$^{210}_{86}$Rn & 6.160  & 3.96  & 7.600 $\times 10^{-5}$& 1.703 $\times 10^{22}$&  1.219$\times 10^{21}$ & 3.957 $\times 10^{-24}$ & 0.0158 \\
%&&&&&& \\
$^{212}_{86}$Rn & 6.380  & 3.17  & 4.686 $\times 10^{-4}$& 1.733 $\times 10^{22}$&  1.237$\times 10^{21}$ & 3.810 $\times 10^{-23}$ & 0.0099 \\
%&&&&&& \\
$^{214}_{86}$Rn & 9.210  & -6.57 & 2.575 $\times 10^{6}$ & 2.084 $\times 10^{22}$&  1.482$\times 10^{21}$ & 4.350 $\times 10^{-14}$ & 0.0399 \\
%&&&&&& \\
$^{216}_{86}$Rn & 8.200  & -4.35 & 1.552 $\times 10^{4}$ & 1.966 $\times 10^{22}$&  1.393$\times 10^{21}$ & 1.011 $\times 10^{-16}$ & 0.1101 \\
%&&&&&& \\
$^{218}_{86}$Rn & 7.260  & -1.45 & 1.854 $\times 10^{1}$ & 1.850 $\times 10^{22}$&  1.307$\times 10^{21}$ & 1.226 $\times 10^{-19}$ & 0.1220 \\
%&&&&&& \\
$^{220}_{86}$Rn & 6.400  & 1.75  & 1.233 $\times 10^{-2}$& 1.736 $\times 10^{22}$&  1.223$\times 10^{21}$ & 6.470 $\times 10^{-23}$ & 0.1558 \\
%&&&&&& \\
$^{222}_{86}$Rn & 5.590  & 5.52  & 2.093 $\times 10^{-6}$& 1.733 $\times 10^{22}$&  1.237$\times 10^{21}$ & 1.055 $\times 10^{-26}$ & 0.1743 \\
%&&&&&& \\
\end{tabular}
\end{ruledtabular}
\end{table*}

\begin{table*}[h]
\label{table2} \caption{Same as table 1, but for the even-even Ra, Th
and U isotopes. }
\begin{ruledtabular}
\begin{tabular}{llllllll}
$Nuclei$& Q$_{\alpha}$ [MeV]&log$_{10}$(T$_{\alpha}$)&$\lambda [ s^{-1} ]$&$v [ fm/s ]$&$\nu_{0} [ s^{-1} ]$ & P &P$_{0}$\\
\hline
%&&&&&& \\
$^{202}_{88}$Ra & 8.020  & -2.58 & 2.666 $\times 10^{2}$ & 1.943 $\times 10^{22}$&  1.411$\times 10^{21}$ & 2.931 $\times 10^{-18}$ & 0.0645 \\
%&&&&&& \\
$^{204}_{88}$Ra & 7.636  & -1.23 & 1.174 $\times 10^{1}$ & 1.896 $\times 10^{22}$&  1.372$\times 10^{21}$ & 1.893 $\times 10^{-19}$ & 0.0452 \\
%&&&&&& \\
$^{206}_{88}$Ra & 7.42   & -0.62 & 2.890 $\times 10^{0}$ & 1.869 $\times 10^{22}$&  1.347$\times 10^{21}$ & 3.750 $\times 10^{-20}$ & 0.0572 \\
%&&&&&& \\
$^{208}_{88}$Ra & 7.27   & 0.15  & 4.907 $\times 10^{-1}$& 1.850 $\times 10^{22}$&  1.329$\times 10^{21}$ & 1.193 $\times 10^{-20}$ & 0.0310 \\
%&&&&&& \\
$^{210}_{88}$Ra & 7.16   & 0.56  & 1.909 $\times 10^{-1}$& 1.836 $\times 10^{22}$&  1.315$\times 10^{21}$ & 5.682 $\times 10^{-21}$ & 0.0256 \\
%&&&&&& \\
$^{212}_{88}$Ra & 7.03   & 1.15  & 4.907 $\times 10^{-2}$& 1.819 $\times 10^{22}$&  1.298$\times 10^{21}$ & 1.982 $\times 10^{-21}$ & 0.0191 \\
%&&&&&& \\
$^{214}_{88}$Ra & 7.27   & 0.40  & 2.759 $\times 10^{-1}$& 1.851 $\times 10^{22}$&  1.316$\times 10^{21}$ & 1.575 $\times 10^{-20}$ & 0.0133 \\
%&&&&&& \\
$^{216}_{88}$Ra & 9.53   & -6.74 & 3.809 $\times 10^{6}$ & 2.120 $\times 10^{22}$&  1.503$\times 10^{21}$ & 5.580 $\times 10^{-14}$ & 0.0454  \\
%&&&&&& \\
$^{218}_{88}$Ra & 8.55   & -4.59 & 2.697 $\times 10^{4}$ & 2.008 $\times 10^{22}$&  1.419$\times 10^{21}$ & 2.049 $\times 10^{-16}$ & 0.0928  \\
%&&&&&& \\
$^{220}_{88}$Ra & 7.60   & -1.74 & 3.809 $\times 10^{1}$ & 1.893 $\times 10^{22}$&  1.333$\times 10^{21}$ & 2.854 $\times 10^{-19}$ & 0.1001  \\
%&&&&&& \\
$^{222}_{88}$Ra & 6.68   & 1.59  & 1.782 $\times 10^{-2}$& 1.774 $\times 10^{22}$&  1.245$\times 10^{21}$ & 1.269 $\times 10^{-22}$ & 0.1127  \\
%&&&&&& \\
$^{224}_{88}$Ra & 5.79   & 5.53  & 2.046 $\times 10^{-6}$& 1.651 $\times 10^{22}$&  1.155$\times 10^{21}$ & 1.216 $\times 10^{-26}$ & 0.1457  \\
%&&&&&& \\
$^{226}_{88}$Ra & 4.87   & 10.73 & 1.291$\times 10^{-11}$& 1.514 $\times 10^{22}$&  1.056$\times 10^{21}$ & 5.861 $\times 10^{-32}$ & 0.2086  \\
%&&&&&& \\
$^{210}_{90}$Th & 8.053  & -1.77 & 4.082 $\times 10^{1}$ & 1.948 $\times 10^{22}$&  1.395$\times 10^{21}$ & 9.009 $\times 10^{-19}$ & 0.0325  \\
%&&&&&& \\
$^{212}_{90}$Th & 7.952  & -1.44 & 1.927 $\times 10^{1}$ & 1.935 $\times 10^{22}$&  1.381$\times 10^{21}$ & 4.616 $\times 10^{-19}$ & 0.0302 \\
%&&&&&& \\
$^{214}_{90}$Th & 7.83   & -1.00 & 6.931 $\times 10^{0}$ & 1.921 $\times 10^{22}$&  1.366$\times 10^{21}$ & 1.978 $\times 10^{-19}$ & 0.0257 \\
%&&&&&& \\
$^{216}_{90}$Th & 8.07   & -1.55 & 2.459 $\times 10^{1}$ & 1.950 $\times 10^{22}$&  1.382$\times 10^{21}$ & 1.256 $\times 10^{-18}$ & 0.0142 \\
%&&&&&& \\
$^{218}_{90}$Th & 9.85   & -7.00 & 6.931 $\times 10^{6}$ & 2.156 $\times 10^{22}$&  1.523$\times 10^{21}$ & 7.388 $\times 10^{-14}$ & 0.0616 \\
%&&&&&& \\
$^{220}_{90}$Th & 8.95   & -5.01 & 7.093 $\times 10^{4}$ & 2.055 $\times 10^{22}$&  1.447$\times 10^{21}$ & 5.102 $\times 10^{-16}$ & 0.0961  \\
%&&&&&& \\
$^{222}_{90}$Th & 8.13   & -2.55 & 2.459 $\times 10^{2}$ & 1.958 $\times 10^{22}$&  1.374$\times 10^{21}$ & 2.514 $\times 10^{-18}$ & 0.0712  \\
%&&&&&& \\
$^{224}_{90}$Th & 7.30   & 0.11  & 5.381 $\times 10^{-1}$& 1.855 $\times 10^{22}$&  1.298$\times 10^{21}$ & 4.512 $\times 10^{-21}$ & 0.0919  \\
%&&&&&& \\
$^{226}_{90}$Th & 6.45   & 3.39  & 2.824 $\times 10^{-4}$& 1.743 $\times 10^{22}$&  1.216$\times 10^{21}$ & 1.901 $\times 10^{-24}$ & 0.1221  \\
%&&&&&& \\
$^{228}_{90}$Th & 5.52   & 7.92  & 8.333 $\times 10^{-9}$& 1.612 $\times 10^{22}$&  1.121$\times 10^{21}$ & 5.626 $\times 10^{-29}$ & 0.1321  \\
%&&&&&& \\
$^{230}_{90}$Th & 4.77   & 12.51 & 2.142 $\times10^{-13}$& 1.498 $\times 10^{22}$&  1.038$\times 10^{21}$ & 1.203 $\times 10^{-33}$ & 0.1714  \\
%&&&&&& \\
$^{232}_{90}$Th & 4.08   & 17.76 & 1.205$\times 10^{-18}$& 1.385 $\times 10^{22}$&  0.957$\times 10^{21}$ & 4.447 $\times 10^{-39}$ & 0.2831  \\
%&&&&&&& \\
$^{218}_{92}$U & 8.773   & -3.29 & 1.358 $\times 10^{3}$  & 2.034 $\times 10^{22}$ & 1.437$\times 10^{21}$  & 3.036 $\times 10^{-17}$ & 0.0311  \\
%&&&&&& \\
$^{220}_{92}$U & 10.30   & -7.22 & 1.156 $\times 10^{7}$  & 2.205 $\times 10^{22}$ & 1.553$\times 10^{21}$  & 1.700 $\times 10^{-13}$ & 0.0438  \\
%&&&&&& \\
$^{224}_{92}$U & 8.620   & -3.15 & 9.904 $\times 10^{2}$  & 2.016 $\times 10^{22}$ & 1.411$\times 10^{21}$  & 1.389 $\times 10^{-17}$ & 0.0505  \\
%&&&&&& \\
$^{226}_{92}$U & 7.701   & -0.30 & 1.386 $\times 10^{0}$  & 1.906 $\times 10^{22}$ & 1.329$\times 10^{21}$  & 1.900 $\times 10^{-20}$ & 0.0549  \\
%&&&&&& \\
$^{228}_{92}$U & 6.80    & 2.76  & 1.205 $\times 10^{-3}$ & 1.790 $\times 10^{22}$ & 1.245$\times 10^{21}$  & 9.088 $\times 10^{-24}$ & 0.1065  \\
%&&&&&& \\
$^{230}_{92}$U & 5.99    & 6.43  & 2.575 $\times 10^{-7}$ & 1.680 $\times 10^{22}$ & 1.164$\times 10^{21}$  & 1.914 $\times 10^{-27}$ & 0.1155  \\
%&&&&&& \\
$^{232}_{92}$U & 5.41    & 9.52  & 2.093 $\times 10^{-10}$& 1.596 $\times 10^{22}$ & 1.103$\times 10^{21}$  & 1.360 $\times 10^{-30}$ & 0.1395  \\
%&&&&&& \\
$^{234}_{92}$U & 4.86    & 13.02 & 6.620 $\times 10^{-14}$& 1.512 $\times 10^{22}$ & 1.042$\times 10^{21}$  & 4.166 $\times 10^{-34}$ & 0.1525  \\
%&&&&&& \\
$^{236}_{92}$U & 4.57    & 14.99 & 7.093 $\times 10^{-16}$& 1.466 $\times 10^{22}$ & 1.007$\times 10^{21}$  & 3.976 $\times 10^{-36}$ & 0.1771  \\
%&&&&&& \\
$^{238}_{92}$U & 4.27    & 17.27 & 3.722 $\times 10^{-18}$& 1.417 $\times 10^{22}$ & 0.970$\times 10^{21}$  & 1.572 $\times 10^{-38}$ & 0.2440  \\
%&&&&&& \\
\end{tabular}
\end{ruledtabular}
\end{table*}

It has been shown that the nuclear structure characteristics can be
detected through the study of the $\alpha$ formation and penetrability
in the above discussions. It is tempting to extend the
study to the superheavy nuclei. Recently, isotopes of the
elements 112, 113, 114, 115, 116 and 118 have been synthesized in
fusion-evaporation reactions using $^{209}$Bi, $^{233,238}$U,
$^{242,244}$Pu, $^{243}$Am, $^{245,248}$Cm and $^{249}$Cf targets
with $^{48}$Ca and $^{70}$Zn beams
\cite{Og05,Hof07,Mo07}. These recent
experimental results have led to new theoretical studies on the
$\alpha$ decay; for example within the relativistic mean field
theory \cite{Sh05}, the DDM3Y interaction \cite{Ba03}, the GLDM
\cite{Zh06,Roy08} and the Skyrme-Hartree-Fock mean field model \cite{Pe07}.
All the above theoretical works are concentrated on the half-lives of these observed superheavy nuclei, and the results are
reasonably consistent with the experimental data. The
characteristics of the $\alpha$ formation and penetrability for the
even-even superheavy nuclei are listed in the table 3. The
preformation factors are distributed between 0.005 and 0.05 showing that the shell effect
structures are also important features for the superheavy
nuclei. The preformation factor of $^{294}118$ is 0.0056, which is
smaller than that of the spherical nucleus $^{210}$Po, implying perhaps 
that the neutron number N=176 is a submagic number. In Ref. \cite{zh04},
it is also pointed out that for the oblate deformed chain of Z$=112$,
the shell closure appears at N$=176$. Thus the preformation of
$\alpha$ is not easy for these heaviest nuclei. The
penetration probabilities P given in the seventh column of the table 3 
always stay between 10$^{-20}$ and
10$^{-15}$. The penetration probabilities are relatively very large,
while the range is narrow. So it is easy for the $\alpha$
particle to escape from these heaviest nuclei as soon as they
form, confirming that the superheavy nuclei are weakly bound.

\begin{table*}[h]
\label{table3} \caption{Same as table 1, but for the even-even
superheavy nuclei. }
\begin{ruledtabular}
\begin{tabular}{llllllll}
$Nuclei$& Q$_{\alpha}$ [MeV]&T$_{\alpha}$&$\lambda$&$v [ fm/s ]$&$\nu_{0} [ s^{-1} ]$ & P &P$_{0}$ \\
\hline
&&&&&&& \\
$^{294}$118 & 11.81 & 0.89 ms & 7.788 $\times 10^{2}$  & 2.365$\times 10^{22}$ & 1.502$\times 10^{21}$  & 9.304 $\times 10^{-17}$ & 0.0056 \\
&&&&&&& \\
$^{292}$116 & 10.66 &  18 ms  & 3.851 $\times 10^{1}$  & 2.246$\times 10^{22}$ & 1.430$\times 10^{21}$  & 5.394 $\times 10^{-19}$ & 0.0499 \\
&&&&&&& \\
$^{290}$116 & 11.00 &  7.1 ms & 9.763 $\times 10^{1}$  & 2.282$\times 10^{22}$ & 1.457$\times 10^{21}$  & 3.813 $\times 10^{-18}$ & 0.0176 \\
&&&&&&& \\
$^{288}$114 & 10.09 & 0.80 s  & 8.664 $\times 10^{-1}$ & 2.185$\times 10^{22}$ & 1.398$\times 10^{21}$  & 5.734 $\times 10^{-20}$ & 0.0108 \\
&&&&&&& \\
$^{286}$114 & 10.33 & 0.13 s  & 5.331 $\times 10^{0}$  & 2.153$\times 10^{22}$ & 1.418$\times 10^{21}$  & 2.476 $\times 10^{-19}$ & 0.0152 \\
&&&&&&& \\
$^{270}$Ds  & 11.20 & 0.1 ms  & 6.931$\times  10^{3}$  & 2.302$\times 10^{22}$ & 1.507$\times 10^{21}$  & 3.486 $\times 10^{-16}$ & 0.0132\\
&&&&&&& \\
$^{266}$Hs  & 10.34 & 2.3 ms  & 3.014 $\times 10^{2}$  & 2.211$\times 10^{22}$ & 1.456$\times 10^{21}$  & 1.134 $\times 10^{-17}$ & 0.0183 \\
&&&&&&& \\
$^{264}$Hs  & 10.58 & 0.25 ms & 2.759 $\times 10^{3}$  & 2.237$\times 10^{22}$ & 1.476$\times 10^{21}$  & 4.143 $\times 10^{-17}$ & 0.0451 \\
&&&&&&& \\
$^{266}$Sg  &  8.76 & 33.88 s & 2.046 $\times 10^{-2}$  & 2.035$\times 10^{22}$ & 1.339$\times 10^{21}$ & 1.473 $\times 10^{-21}$ & 0.0104 \\
&&&&&&& \\
$^{260}$Sg  &  9.93 & 8.51 ms & 8.144 $\times 10^{1}$  &  2.167$\times 10^{22}$ & 1.438$\times 10^{21}$ & 3.359 $\times 10^{-18}$ & 0.0169 \\
\end{tabular}
\end{ruledtabular}
\end{table*}

\begin{figure*}[t]
%\centering
\includegraphics[scale=0.8]{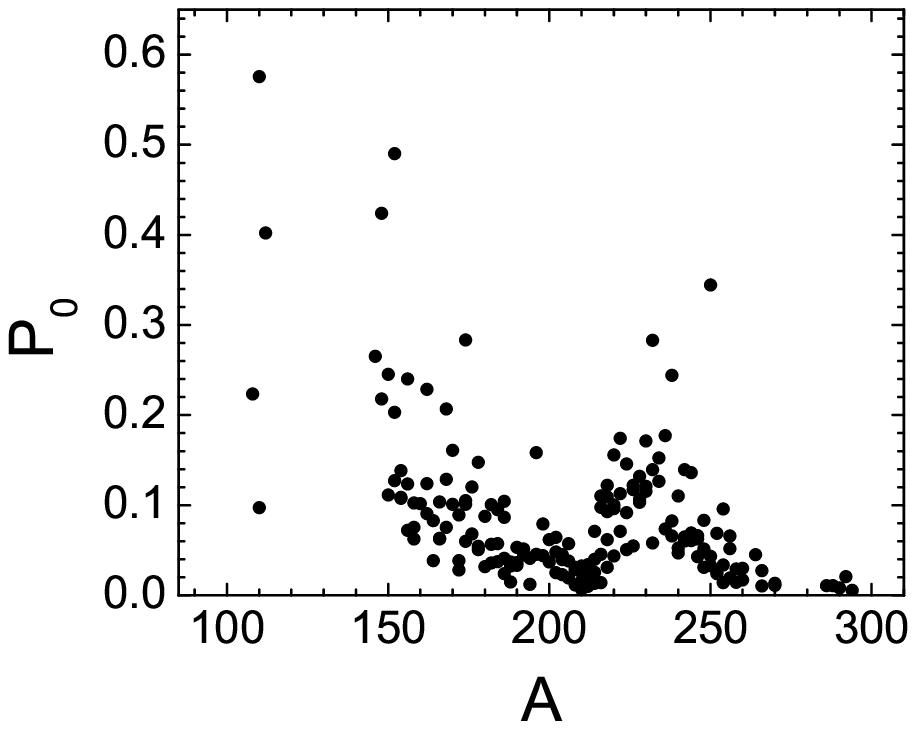}
\caption{$\alpha$ preformation factors for 180 even-even nuclei
(Z=52-118).} \label{Fig6}
\end{figure*}

The experimental but model-dependent preformation factors of the 180
even-even nuclei from Z=52 to Z=118 are displayed in Fig. \ref{Fig6}. In general,
the formation factor decreases with the increase
of the mass number A. It can be understand easily that the
theoretical $\alpha$ decay GLDM half-lives are very slightly higher
than the experimental data for the light nuclei and lower for the heaviest systems, when the preformation factors are
neglected. The preformation factors decrease with
mass number till about 210, then, the preformation
factors jump nearly one order of magnitude, and finally
decrease again, due to the neutron number N=126 in this mass
region.

\section{Conclusion and outlook}
A global study of the $\alpha$ formation and
penetration is presented for the even-even nuclei from Z=52 to Z=118. The decay
constants are obtained from the experimental half-lives. The
penetration probabilities are calculated from the WKB approximation
and the penetration barriers are constructed with the GLDM. After using
a classical method to estimate the assault frequencies the
preformation factors are extracted systematically. The
preformation factors are not constant for these nuclei,
and show regular law for the studied isotopes. Clearly the closed shell structures play
the key role for the preformation mechanism, and the more the nucleon
number is close to the magic nucleon numbers, the more the
preformation of $\alpha$ cluster is difficult inside the mother nucleus. The decay constant is the product of the
preformation, the penetration, and the assault frequency. The penetrability determines mainly the behaviour of the decay constant. The study is extended to
the even-even superheavy nuclei. The penetration
probabilities are relatively very large and its range is
narrow for the heaviest nuclei. So it is easy for the $\alpha$
particle to escape from these nuclei as soon as they are formed,
confirming that the superheavy nuclei are weakly bound. Thus, it is interesting and timely to study the preformation factors
 microscopically and
systematically. It is also desired to extend the study to the odd
A and odd-odd nuclei and extract possibly analytical formulas of
preformation factors, which should be functions of the nucleon
numbers and the $\alpha$ decay energy. It would be also important to take into account the deformations, which may change
the results by perhaps one order of magnitude.

\section{Acknowledgments}
H.F. Zhang is thankful to Prof. Bao-Qiu Chen, Zhong-Zhou Ren,
En-Guang Zhao for valuable discussions. The work of H.F. Zhang was
supported by the National Natural Science Foundation of China
(10775061,10505016,10575119,10175074), the Knowledge Innovative
Project of CAS (KJCX3-SYW-N2), the Major Prophase Research Project
of Fundamental Research of the Ministry of Science and Technology
of China (2007CB815004).

%\begin{references}

\end{document}